# Effect of alloying and microstructure on formability of advanced high-strength steels processed via quenching and partitioning

P. Xia[1,2], F. Vercruysse[3], C. Celada-Casero[4], P. Verleysen[3], R.H. Petrov[3,4], I. Sabirov[1] *,

J.M. Molina-Aldareguia[1], A. Smith[5], B. Linke[6], R. Thiessen[6], D. Frometa[7], S. Parareda[7],

A. Lara[7]

[1] IMDEA Materials Institute, 28906 Getafe, Spain

[2] Currently at State Key Laboratory of Metal Matrix Composites, Shanghai Jiao Tong University, 200240 Shanghai, China

[3] Department of Electromechanical, Systems & Metal Engineering, Research Group Materials Science and Technology, Ghent University, Tech Lane Science Park Campus A 46, Zwijnaarde, Belgium

[4] Department of Materials Science and Engineering, Delft University of Technology, 2628 CD Delft, The Netherlands

[5] RINA Consulting – Centro Sviluppo Materiali Spa, 00128 Rome, Italy

[6] ThyssenKrupp Steel Europe AG, 47166 Duisburg, Germany

[7] Eurecat, Centre Tecnològic de Catalunya, 08243 Manresa, Spain

**Abstract**

The article focuses on the effect of alloying and microstructure on formability of advanced high strength steels (AHSSs) processed via quenching and partitioning (Q&P). Three different Q&P steels with different combination of alloying elements and volume fraction of retained austenite are subjected to uniaxial tensile and Nakajima testing. Tensile mechanical properties are determined, and the forming limit diagrams (FLDs) are plotted. Microstructure of the tested samples is analyzed, and dramatic reduction of retained austenite fraction is detected. It is demonstrated that all steels are able to accumulate much higher amount of plastic strain when tested using Nakajima method. The observed phenomenon is related to the multiaxial stress state and strain gradients

---

* Corresponding author.
E-mail: ilchat.sabirov@imdea.org
Postal address: IMDEA Materials Institute, Calle Eric Kandel 2, 28906 Getafe, Spain

through the sheet thickness resulting in a fast transformation of retained austenite, as well as the ability of the tempered martensitic matrix to accumulate plastic strain. Surprisingly, a Q&P steel with the highest volume fraction of retained austenite and highest tensile ductility shows the lowest formability among studied grades. The latter observation is related to the highest sum of fractions of initial fresh martensite and stress/strain induced martensite promoting formation of microcracks. Their role and ability of tempered martensitic matrix to accumulate plastic deformation during forming of Q&P steels is discussed.

**Keywords**: advanced high strength steels, quenching and partitioning, strength, ductility, formability, retained austenite

# 1. Introduction

## 1.1 Quenched and partitioned steels as attractive candidates for automotive applications

The development of materials for modern automotive industry is defined by numerous and often contradictory constraints including manufacturing capabilities of the industrial processing lines, high mechanical strength, lightweighting of parts and structural components, sufficient formability for metalforming operations and many others [1]. The steel manufacturers have continuously been presenting innovative solutions to the automotive industry, since steels have the ability to adapt to the changing requirements [2-5]. Over the past decades, significant research efforts have been directed towards the development of several types of advanced high strength steels (AHSSs) [3], since they provide an opportunity for the manufacturing of cost-effective, light-weight parts with improved safety and optimized environmental performance [4,6,7].

AHSSs are very complex materials with carefully selected chemical compositions and multiphase microstructures resulting from precisely controlled heating and cooling processes [1,8]. Their classical development is usually based on specific combinations of alloying elements and/or controlling the thermo-mechanical and thermal treatment parameters. Three generations of AHSSs have been developed up-to-date [9]. The first generation AHSS comprises complex phase (CP), dual phase (DP), and transformation induced plasticity (TRIP) steels. They show high mechanical strength but their tensile ductility is relatively low. The second generation AHSS consists of highly alloyed austenitic steels, including stainless steels and twinning induced plasticity (TWIP) steels. They demonstrate a combination of high tensile strength and very high tensile ductility. Their superior mechanical properties make them very attractive for engineering, but their industrial application is limited by their very high cost and complexity of industrial processing. Over the last two decades, significant research efforts have been devoted to the development of the third generation AHSS. In 2003, Speer *et al*. [10] firstly proposed a new concept for processing of AHSSs containing retained austenite, which belong to the third generation. The idea was based on a new understanding of carbon diffusion from martensite into retained austenite for its stabilization, the so-called "quenching and partitioning" (Q&P) process.

The Q&P process consists of a two-step heat treatment. After heating to generate a fully austenitic or partially austenitic (austenite and ferrite) microstructure, the steel is quenched to a pre-defined temperature below the martensite start ($M_s$) but above the martensite finish ($M_f$) temperatures [11,12]. The microstructure at this quenching temperature (QT) consists of martensite and untransformed austenite, if it was fully austenitized, or contains additionally ferrite, if it was intercritically annealed. The second step consists of either isothermal holding at the QT or at a higher partitioning temperature (PT). The aim of the latter step is to enrich with carbon the untransformed austenite remaining at the QT through carbon depletion of the carbon supersaturated martensite. The martensite loses carbon by diffusion into the retained austenite but also by tempering. Part of the retained austenite may not be stabilized, and during final quenching to room temperature it can transform to fresh, untempered martensite. A microstructure with martensite matrix (consisting mainly of tempered martensite with minor fraction of fresh martensite) and metastable retained austenite is obtained. Ferrite is additionally present in the microstructure, if the steel is annealed at the intercritical temperature [11,12]. It should be noted that the presence of ferrite in an intercritically annealed steel results in higher carbon content in austenite (and in primary martensite formed from it). These in turn affect the carbon partitioning process thus modifying morphology and improving stability of retained austenite [13,14].

Numerous investigations have clearly demonstrated that Q&P processed steels show a good combination of tensile strength and ductility. A prerequisite, however, is that the optimal microstructure is obtained by imposing the most appropriate Q&P treatment to a steel with well-determined chemistry [12]. The principles of microstructural design in Q&P steels for improvement of their mechanical properties have been understood to a satisfactory level. However, the steel performance for automotive applications is not governed just by its mechanical strength and ductility under uniaxial tension. Enhanced mechanical properties need to be combined with satisfactory application-related properties (such as formability, weldability, energy absorption in crash, etc.), which have been very scarcely investigated for the Q&P steels. Large-scale commercialization of Q&P steels is currently limited by the lack of knowledge on their application-related properties. Present work focuses on one of their most important application-related properties, i.e., formability.

*1.2 Formability of AHSS*

Formability is generally defined as the capability of a sheet metal to undergo plastic deformation to a given shape without formation of defects [15]. The formability of first and second generation AHSSs has already been thoroughly studied. The outcomes of these studies can be roughly summarized as follows. The second generation AHSSs shows superior formability compared to the grades of the first generation [9]. In the first generation, TRIP steel performs better compared to CP steel, which is followed by DP steel [16]. The better formability of TRIP steel is related to its higher strain hardening ability [17]. The maximum admissible limiting strains strongly depend on several physical factors, including strain rate sensitivity, work-hardening and plastic anisotropy induced by cold rolling processes [17]. Analysis of formability of various DP, CP and TRIP steels showed that both strength and crack growth resistance were the key factors determining the formability of the materials [18,19]. In the second generation, TWIP steels show better formability compared to the stainless steels [20]. Combination of TRIP and TWIP effects in steels with high Mn content leads to high strength and formability [21].

The research activities on formability of Q&P processed steels have been very limited and just a few relevant publications can be found in the current literature. The Q&P process, carried out in research labs using dilatometers or Gleeble thermo-mechanical simulators, allows to produce only small samples which cannot be used for formability testing. In [22], it was demonstrated that a TRIP steel showed a higher formability compared to a Q&P steel. In [23,24], quenching-partitioning-tempering (Q-P-T) heat treatment applied to a Fe-0.20C-1.49Mn-1.52Si-0.58Cr-0.05Nb (wt. %) steel led to a better formability compared to the conventional quenching-tempering (Q-T) treatment. This observation was related to a better strain hardening ability of the Q-P-T treated steel due to the higher volume fraction of retained austenite compared to the Q-T treated counterpart. The yield ratio of Q-P-T samples (0.81) was lower compared to that of the Q-T samples (0.92). In [25], the stretch-flangeability and tensile ductility of a Fe-0.2C-2.3Mn-1.4Si steel after different Q&P treatments was assessed and compared to those after conventional heat treatments, including quenching and tempering (Q&T) and quenching and austempering (QAT). It was shown that the Q&T treatment, resulting in a tempered martensitic microstructure, leads to the highest hole expansion capacity (76 %) and yield ratio (0.9), but the lowest tensile uniform ductility (2.8 %). The QAT treated alloy demonstrated limited stretch-flangeability (32 %) due to the fresh martensite islands present in the bainitic microstructure, despite the good tensile uniform ductility of the material (6.8 %) and low yield ratio (0.68). The Q&P microstructures showed a good combination between stretch-flangeability and tensile ductility, although their performance depended on the applied Q&P cycles. No thorough systematic studies on the effect of microstructure on formability of Q&P processed steels have been carried out up-to-date. The role of retained austenite in plastic deformation during formability testing is yet to be explored. The main objective of the present work is to study the effect of the microstructure on the formability of three AHSSs with varying chemical compositions processed via Q&P route. Variations in chemical compositions allow to produce three different microstructures with varying volume fractions of retained austenite.

**2. Materials and experimental procedures**

*2.1. Materials and processing*

The chemical compositions of the studied steel grades are presented in Table 1. Based on their chemical compositions, the steels are further referred to as 'base', 'base+Cr' and 'base+Cr+Nb'. Their $M_S$ temperatures measured by dilatometry after full austenitization and quenching at 50 °C/s are also provided in Table 1.

Table 1 Chemical composition (in wt. %) of the studied steel grades and their $M_S$ temperatures.

| Grade | C | Si | Mn | Al | Cr | Nb | $M_S$ (°C) |
|---|---|---|---|---|---|---|---|
| base | 0.2 | 1.25 | 2.4 | 0.02 | 0 | 0 | 389 |
| base+Cr | 0.2 | 1.25 | 2.4 | 0.02 | 0.3 | 0 | 388 |
| base+Cr+Nb | 0.2 | 1.25 | 2.4 | 0.02 | 0.3 | 0.025 | 379 |

The cold rolled 1.5 mm thick sheets were subjected to Q&P treatments on an industrial MULTIPAS annealing simulator VATRON (Vatron GmbH, Austria). It is based on resistance heating and allows annealing of steel sheet with rapid heating (up to 100 K/s) and cooling (up to 70 K/s) rates.

All alloys were fully austenitized at 870 ºC for 100 s and quenched to a quenching temperature (QT) in order to obtain microstructures consisting of martensite and austenite. The samples were then reheated to a partitioning temperature (PT) of 400 ºC and kept isothermally for a partitioning time (Pt) of 50 s followed by final quenching to room temperature. These partitioning conditions ensure a considerable partitioning of carbon from martensite to austenite, avoiding significant tempering of the martensite [26,27]. Table 2 outlines the main parameters of the Q&P treatment applied to different alloys. The quenching temperatures were selected based on preliminary dilatometry data and microstructure characterization so that the stabilization of a sufficiently high volume fraction of austenite (around 10 %) is ensured within a processing window of QT ±20 °C.

Table 2 Quenching and partitioning (Q&P) parameters.

| Grade | Austenitization temperature [ºC] | Austenitization time [s] | Quenching temperature QT, [ºC] | Partitioning temperature, PT [ºC] | Partitioning time, Pt [s] |
|---|---|---|---|---|---|
| base | 870 | 100 | 300 | 400 | 50 |
| base+Cr | 870 | 100 | 280 | 400 | 50 |
| base+Cr+Nb | 870 | 100 | 290 | 400 | 50 |

## 2.2 Mechanical testing

### 2.2.1 Uniaxial tensile testing

Uniaxial tensile tests were carried out for all Q&P treated steels according to DIN EN ISO 6892-1 Standard. For each studied alloy, five tensile specimens were taken transverse to the rolling direction and machined to A50 samples. The engineering stress – engineering strain curves were plotted, and the offset yield strength, $\sigma_{0.2}$, ultimate tensile strength, $\sigma_{UTS}$, the uniform elongation, $\varepsilon_u$, and the elongation to failure, $\varepsilon_f$, were determined from the stress-strain curves. Strain hardening curves up to uniform elongation were generated from the true stress – true strain curves to analyze strain hardening behavior of the materials.

In order to assess the planar anisotropy, additional tensile tests were performed in the rolling, 45° and transverse direction on miniaturised samples following the procedures and measurement techniques described in [28]. The Lankford coefficient R was obtained using the following equation [29]:

$$R = \frac{\epsilon^{w,pl}}{\epsilon^{th,pl}} = \frac{\epsilon^{w,pl}}{-(\epsilon^{w,pl} + \epsilon^{l,pl})}$$

with $\epsilon^{l,pl}$, $\epsilon^{th,pl}$ and $\epsilon^{w,pl}$ the true plastic strain components in respectively the tensile, thickness and width direction of the sample at the end of the uniform elongation. $R_{av}$ and $\Delta R$ were respectively derived as:

$$R_{av} = \frac{R_{0°} + 2R_{45°} + R_{90°}}{4}$$

$$\Delta R = \frac{R_{0°} - 2R_{45°} + R_{90°}}{2}$$

in which $R_{0°}$, $R_{45°}$ and $R_{90°}$ denote the R values obtained in the 0°, 45° and 90° directions with respect to the rolling direction.

*2.2.2 Nakajima testing*

The forming and fracture limit curves are established by experiments that provide limits for pairs of the in-plane, principal strains, i.e., $\varepsilon_{minor}$ and $\varepsilon_{major}$, obtained for various, proportional deformation paths. In order to determine the limit curves, strain paths ranging from pure shear ($\varepsilon_{major} = -\varepsilon_{minor}$) to equi-biaxial tension ($\varepsilon_{major} = \varepsilon_{minor}$) are needed. However, in practice, shear is not reached in the blank holder region and, very often, tension ($\varepsilon_{major} = -2\varepsilon_{minor}$ for isotropic materials) is used as the left limit in the FLD [15]. One of the most common methods for measuring sheet forming limits is the Nakajima test. It is recommended by the ISO 12004 standard [15,30]. The Nakajima test consists of drawing rectangular specimens having different widths clamped in a circular die with a hemispherical punch [15]. By varying the width of the specimen, both the positive and negative minor strain domains of the FLD are covered [15]. In the tests, the sheet sample is clamped between the upper die and the lower blank holder, while the hemispherical punch stretches the material until failure. The initial strain in the center of the sample is always bi-axial. Then gradually the strain path changes towards the final path imposed by the sample geometry and the material properties [15,31].

Due to the size of the manufactured Q&P steel sheets, miniaturized Nakajima tests were carried out to derive the formability limits of the materials. Square and rectangular specimens with varying dimensions were cut with the axis perpendicular to the rolling direction of the sheet material. The geometries of specimens are presented in Fig. 1. Tests were carried out in a double acting hydraulic press with 1500 and 500 kN drawing capacity on the top and bottom pistons, respectively. The bottom piston is used to clamp the sample, while the top piston is used to impose the punch deformation to deform the samples. The punch used in the tests has a diameter of 50 mm. The press-holder is equipped with a circular rib to avoid material sliding. All tests were performed with a punch speed of 1 mm/s.

Reduction of the friction between the punch and the sample is crucial for the reliability of the test outcome. To this purpose, a stack composed of grease and polymers foils was used between the specimen and the punch, composed of vaseline grease, circular blank of 0.1 mm thick PTFE foil, vaseline grease, circular blank of 2 mm thick soft PVC foil, vaseline grease, circular blank of 0.1 mm thick PTFE foil and vaseline grease. The diameter of the foil blanks was 50 mm. In this way, fractures within a distance less than 15% of the punch diameter away from the apex of the dome were obtained.

Full field strain measurements were done using the digital image correlation (DIC) system ARAMIS developed by GOM GmbH. The system is based on two CCD cameras monitoring the surface during formability testing. A stochastic speckle pattern was applied by spray painting on the sample surface before testing to follow the local deformation [32]. The camera frame rate was 10 fps.

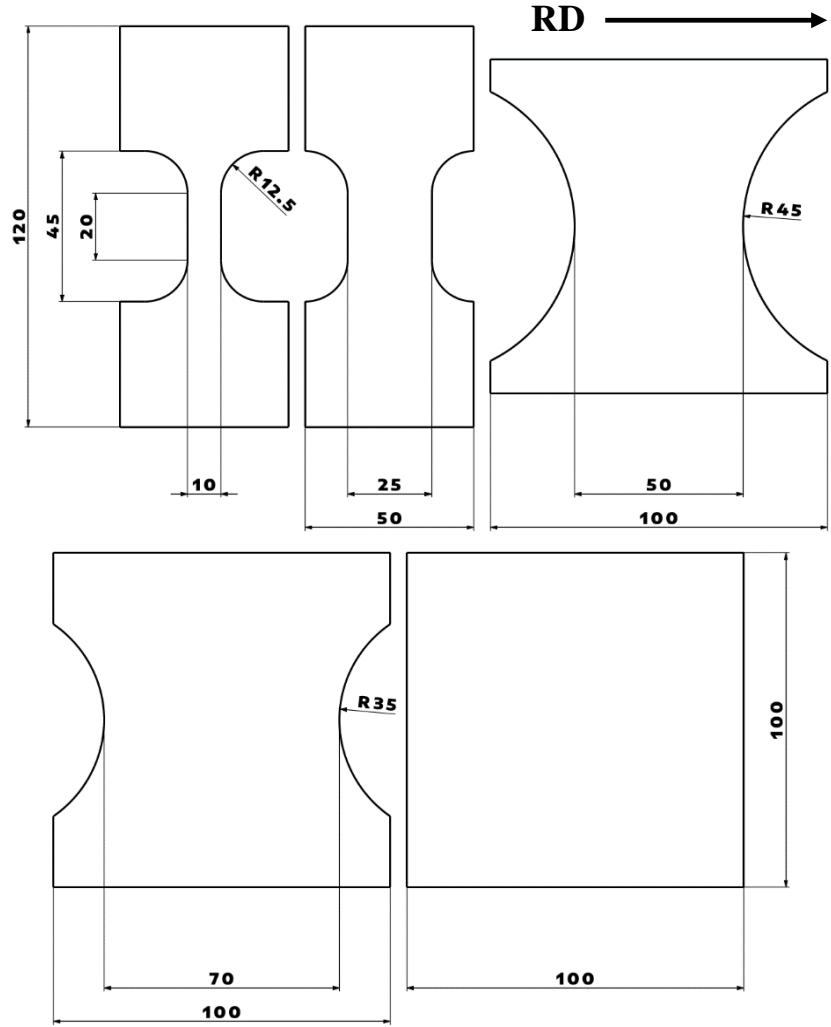

Fig. 1 Geometries of the specimens tested to determine FLC. The specimens from left to right and from top to bottom are denoted as A010, A025, A050, A070 and A100 referring to the width of the gauge section.

The evaluation of forming limits was performed following the recommendations of the ISO12004 standard [30]. Three sections perpendicular to the fracture plane were chosen in the middle section of the specimens, as indicated in Fig. 2. The sections were separated from each other by a distance of 1 mm. The image before fracture was used for the determination of the forming and fracture limit strains using the values of major and minor true strain ($\varepsilon_{major}$, $\varepsilon_{minor}$) obtained by DIC along the length of the three sections.

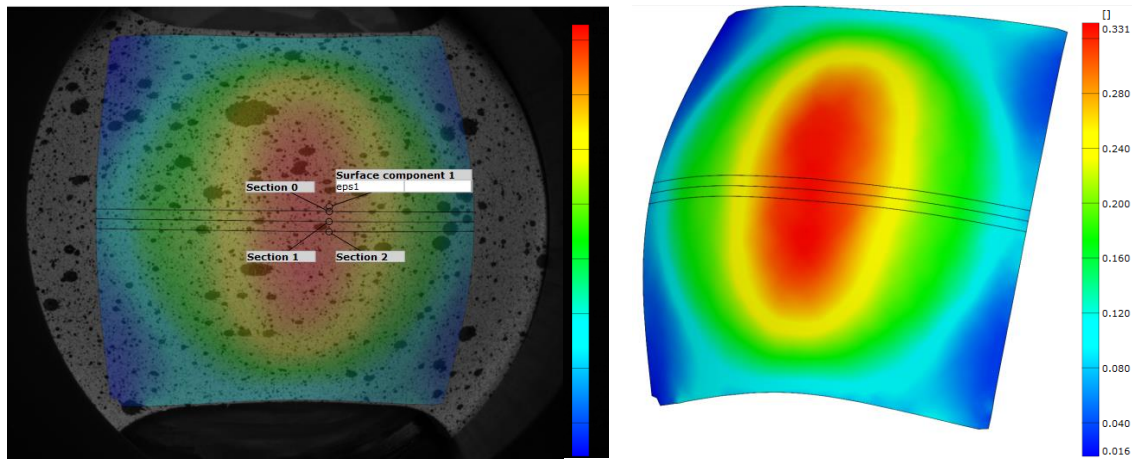

Fig. 2 Sections perpendicular to the fracture plane used to evaluate the forming limit.

*2.3 Microstructural characterization*

Specimens for electron backscatter diffraction (EBSD) microstructural characterization were ground and polished to a mirror-like surface using standard metallographic techniques and final polishing with OPS for 10 minutes. Electron backscatter diffraction (EBSD) analysis at high magnifications was performed using a FEI Quanta™ Helios NanoLab 600i, equipped with a NordlysNano detector controlled by the AZtec Oxford Instruments Nanoanalysis (version 2.4®) software. The data were acquired at an accelerating voltage of 18 kV, a working distance of 8 mm, a tilt angle of 70° and a step size of 50 nm. The orientation data were post-processed with HKL post-processing Oxford Instruments Nanotechnology (version 5.1©) software. All kernel average misorientation (KAM) maps were calculated for the $3^{rd}$ nearest neighbor and the upper limit was 5°. EBSD analysis at low magnifications was carried out using an AMETEK EDAX-TSL EBSD system coupled with the Quanta 450 FEG-SEM. The orientation data were post-processed by means of AMETEK EDAX-TSL-OIM Data analysis (version 7.3) after applying the "grain confidence index standardization" and "grain dilation" clean-up procedures in one step to the original orientation data. Thus obtained data were used to estimate the prior austenite grain size by using the ARPGE software [33]. The latter allows to automatically reconstruct the parent grains from EBSD data of materials constituted only of daughter grains formed in a displacive manner by applying specific "transformation law". In the current case, we used Kurdjumov-Sachs (K-S) orientation relationship between the parent austenite and product martensite phases [34]. To derive the average grain size of the retained austenite and martensite, only grain boundaries with misorientation angles larger than 5° and grains consisting of at least 3 pixels were considered.

Since EBSD measurements do not allow to resolve nanoscale features having a size less than the pixel size (i.e. film type retained austenite) [36], XRD measurements were additionally carried out to estimate the total volume fraction of retained austenite. A Siemens Kristalloflex D5000 diffractometer operating with a Mo-$K_\alpha$ source at 40 kV and 40 mA was employed. The retained austenite fraction was determined based on the $\{200\}\alpha$, $\{211\}\alpha$, $\{220\}\gamma$ and $\{311\}\gamma$ diffraction peaks after subtracting the $K_{\alpha 2}$ and the background radiation from the raw data [36]. The used Mo-$K_\alpha$ source can resolve lattices with a spacing of 0.036 nm. The detectable size and volume fraction of retained austenite are in the range of 10 Å and 0.5 %, respectively [37].

## 3. Results and discussion

### *3.1 Mechanical behaviour of the Q&P steel grades*

*3.1.1 Tensile properties*

Table 3 summarizes the mechanical properties measured on the Q&P processed steel grades. The results can be summarized as follows. The $\sigma_{0.2}$ values are superior to 930 MPa in all samples. The highest $\sigma_{0.2}$ value of 1003 MPa is reached in the base+Cr sample (Table 3). The ultimate tensile strength of all Q&P treated grades exceeds 1200 MPa. The highest $\sigma_{UTS}$ value of 1260 MPa is demonstrated by the base+Cr+Nb sample. The average values of both uniform tensile elongation $\varepsilon_u$ and elongation to failure $\varepsilon_f$ tend to increase with increasing alloying level, though this increase is more pronounced in uniform elongation than in elongation to failure.

Table 3  Tensile properties of the studied steels.

| Grade | $\sigma_{0.2}$ *[MPa]* | $\sigma_{UTS}$ *[MPa]* | $\varepsilon_u$ *[%]* | $\varepsilon_f$ *[%]* |
|---|---|---|---|---|
| base | 988 ± 16 | 1203 ± 9 | 6.3 ± 0.2 | 10.8 ± 0.3 |
| base+Cr | 1003 ± 15 | 1229 ± 13 | 7.4 ± 0.9 | 11.4 ± 0.7 |
| base+Cr+Nb | 936 ± 26 | 1260 ± 4 | 8.7 ± 0.4 | 11.8 ± 0.7 |

Fig. 3 illustrates strain hardening curves derived from the true stress–true strain curves. Five curves are presented for each material for statistical reasons. It can be seen that the strain hardening ability of the grades also improves with the increasing alloying content. The strain hardening behavior of the base+Cr grade scatters widely (particularly in uniform elongation with values from 6.5 to 8.3 %). The base+Cr+Nb grade shows the highest overall strain hardening with remarkably little scatter towards the end. There is a clear correlation between strain hardening ability and uniform elongation in the studied steels.

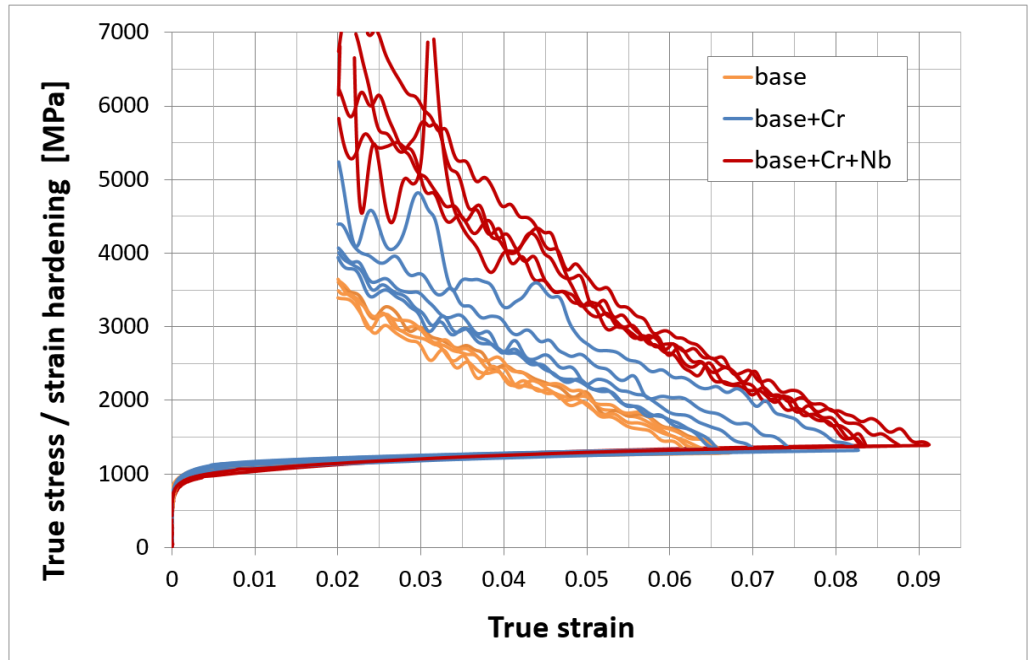

Fig. 3 Strain hardening and true stress – true strain curves from different tests of the studied Q&P processed grades.

*3.1.2 Formability of the Q&P grades*

Table 4 gives the $R_{av}$ and $\Delta R$ values obtained for the three steel grades. Average Lankford coefficients well above 1 are found for all steels, indicating a good forming potential. The relatively low $\Delta R$ values show a limited planar anisotropy.

Table 4  Average $R_{av}$ and $\Delta R$ Lankford values.

| Grade | $R_{av}$ [-] | $\Delta R$ [-] |
|---|---|---|
| base | 1.53 | 0.05 |
| base+Cr | 1.54 | 0.04 |
| base+Cr+Nb | 1.45 | 0.08 |

Fig. 4 demonstrates typical appearance of the samples after Nakajima testing. It can be seen that all tested samples cracked in the top of the dome, as required by the ISO standards.

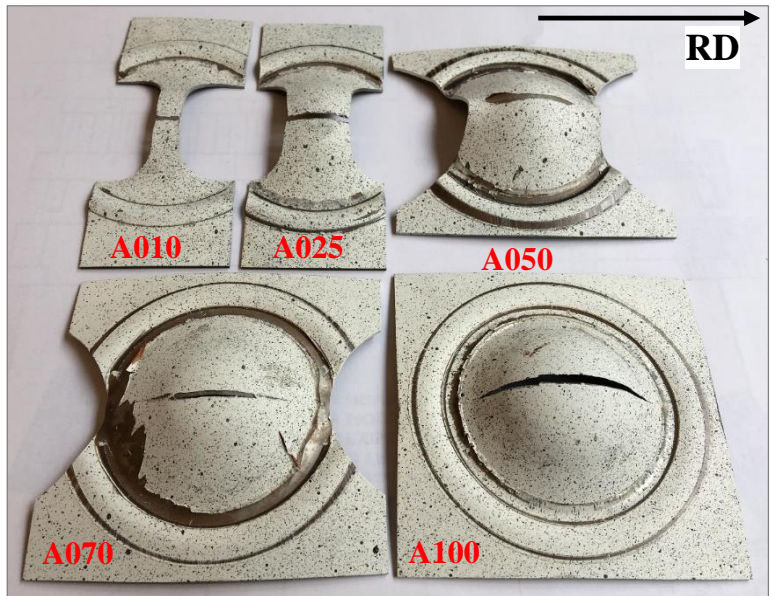

Fig. 4 Typical appearance of samples after miniaturized Nakajima testing. From left to right from top to bottom: A010, A025, A050, A070, A100.

The forming and fracture limit curves (see Section 2.2.2) are presented in Fig. 5. Exact values of the limit major and minor strains are summarized in Table 5.

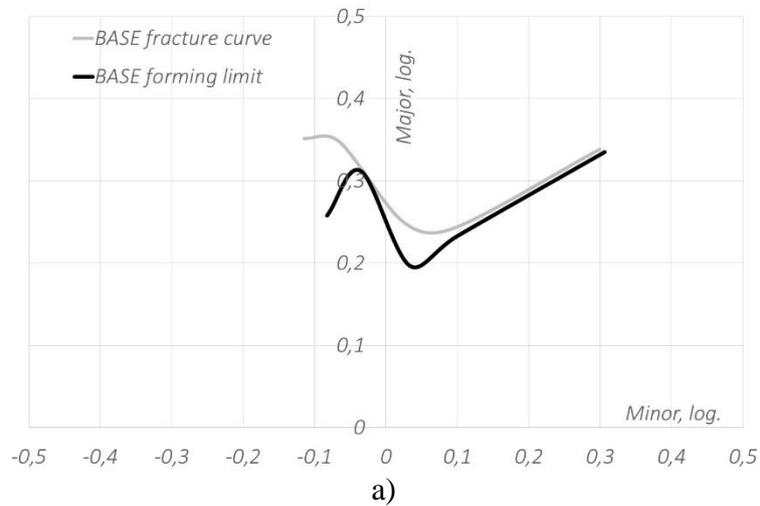

a)

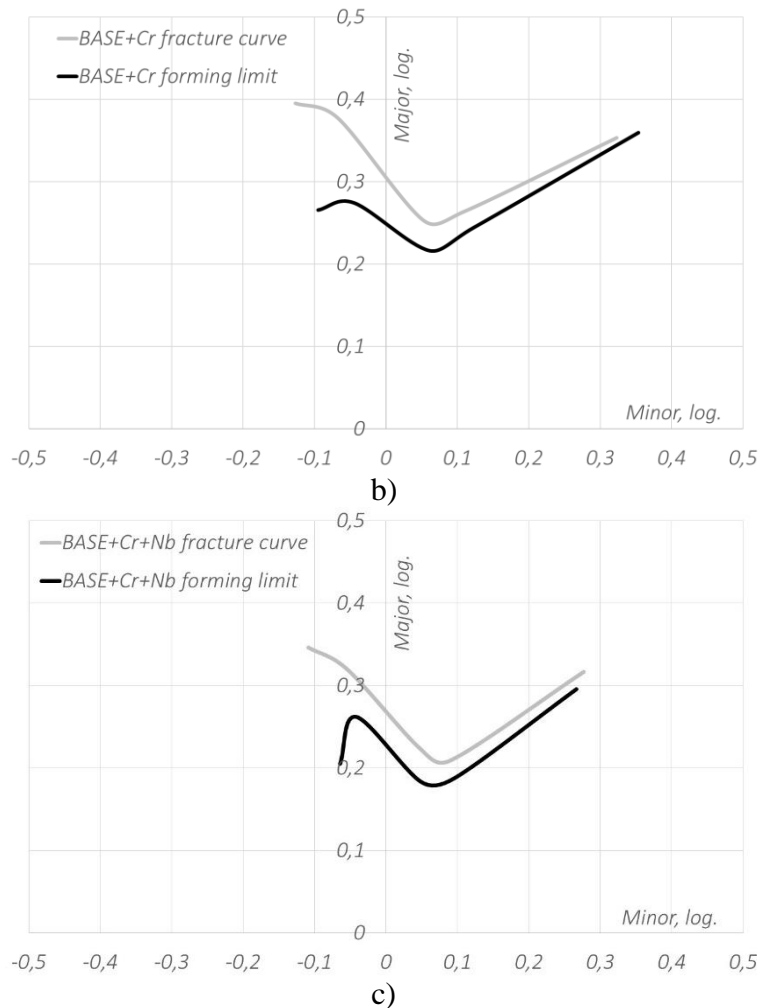

Fig. 5 Forming and fracture limit curves obtained from fracture points and ISO12004 standard fitting for the steels: a) base; b) base+Cr and c) base+Cr+Nb.

Table 5 Major and minor strains obtained from analysis of the forming and fracture limit curves.

| Steel | Geometry | Fracture | | ISO12004 | |
|---|---|---|---|---|---|
| | | Maj [log] | Min [log] | Maj [log] | Min [log] |
| Base | A010 | 0.381 | -0.117 | 0.249 | -0.103 |
| | A025 | 0.397 | -0.072 | 0.300 | 0.002 |
| | A050 | 0.254 | 0.044 | 0.219 | 0.054 |
| | A070 | 0.256 | 0.088 | 0.235 | 0.101 |
| | A100 | 0.375 | 0.312 | 0.378 | 0.383 |
| Base+Cr | A010 | 0.395 | -0.126 | 0.265 | -0.095 |
| | A025 | 0.377 | -0.067 | 0.274 | -0.044 |
| | A050 | 0.253 | 0.053 | 0.216 | 0.061 |
| | A070 | 0.264 | 0.110 | 0.244 | 0.123 |
| | A100 | 0.353 | 0.323 | 0.359 | 0.354 |
| Base+Cr+Nb | A010 | 0.346 | -0.108 | 0.205 | -0.064 |
| | A025 | 0.319 | -0.053 | 0.262 | -0.041 |
| | A050 | 0.224 | 0.048 | 0.183 | 0.049 |
| | A070 | 0.209 | 0.090 | 0.188 | 0.097 |
| | A100 | 0.317 | 0.277 | 0.296 | 0.266 |

In order to compare the different steel grades, Fig. 6 gives the fracture limit curves of all studied grades together with the curve of a TRIP700 steel for comparison. It can be seen that the base alloy shows formability values similar to the base+Cr alloy, except for equi-biaxial stretching conditions for which the base alloy shows a somewhat better performance. Microstructural changes induced by additional microalloying by Nb noticeably reduce the formability of the Q&P treated steel. The fracture limits of the base+Cr+Nb alloy lay well below the curves for the other steels for all deformation conditions. The TRIP700, on the contrary, outperforms the Q&P steel grades.

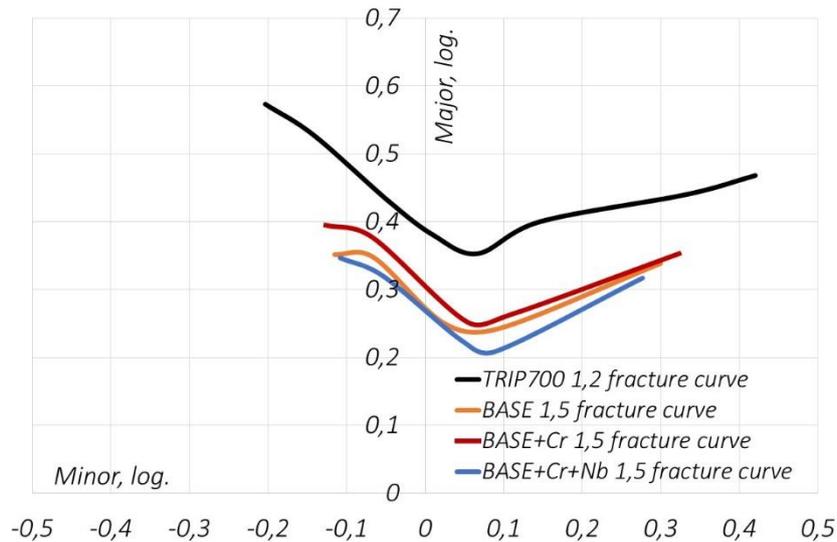

Fig. 6 Comparison of fracture FLDs of the studied grades.

*3.2 Microstructure of the Q&P treated grades before and after Nakajima testing*

*3.2.1 Initial microstructure of the Q&P treated steels*

Band contrast EBSD maps overlaid by phase maps, where fcc phase is in green, are presented in Fig. 7a,c,e. All studied Q&P treated samples showed a homogeneous microstructure of similar morphology, where the matrix consists of irregular-shaped blocks of tempered martensite. The relevant KAM maps (Fig. 7b,d,f) illustrate the matrix lath martensite structure with lath boundaries characterized by higher KAM values of 2-2.5°. The retained austenite grains are homogeneously dispersed in the matrix. Both blocky and large interlath retained austenite grains are visible on the EBSD maps (Fig. 7a,c,e). Due to a more distorted lattice, fresh martensite formed during the final quench to room temperature shows up as the dark grey areas, which can be found in the vicinity of the green retained austenite (Fig. 7a,c,e). On the KAM maps (Fig. 7b,d,f), they appear as non-indexed areas (white pixels).

The volume fraction of retained austenite measured by EBSD tends to grow with the alloying content from 7.0 % in the base steel to 8.0 % in the base+Cr steel to 9.9 % in the base+Cr+Nb steel (Table 6). The average grain size of retained austenite seen on the EBSD maps of all analyzed samples corresponds to an equivalent grain diameter of 0.5 µm. Very low volume fractions of fresh martensite (1 – 1.9 %) were obtained in all alloys.

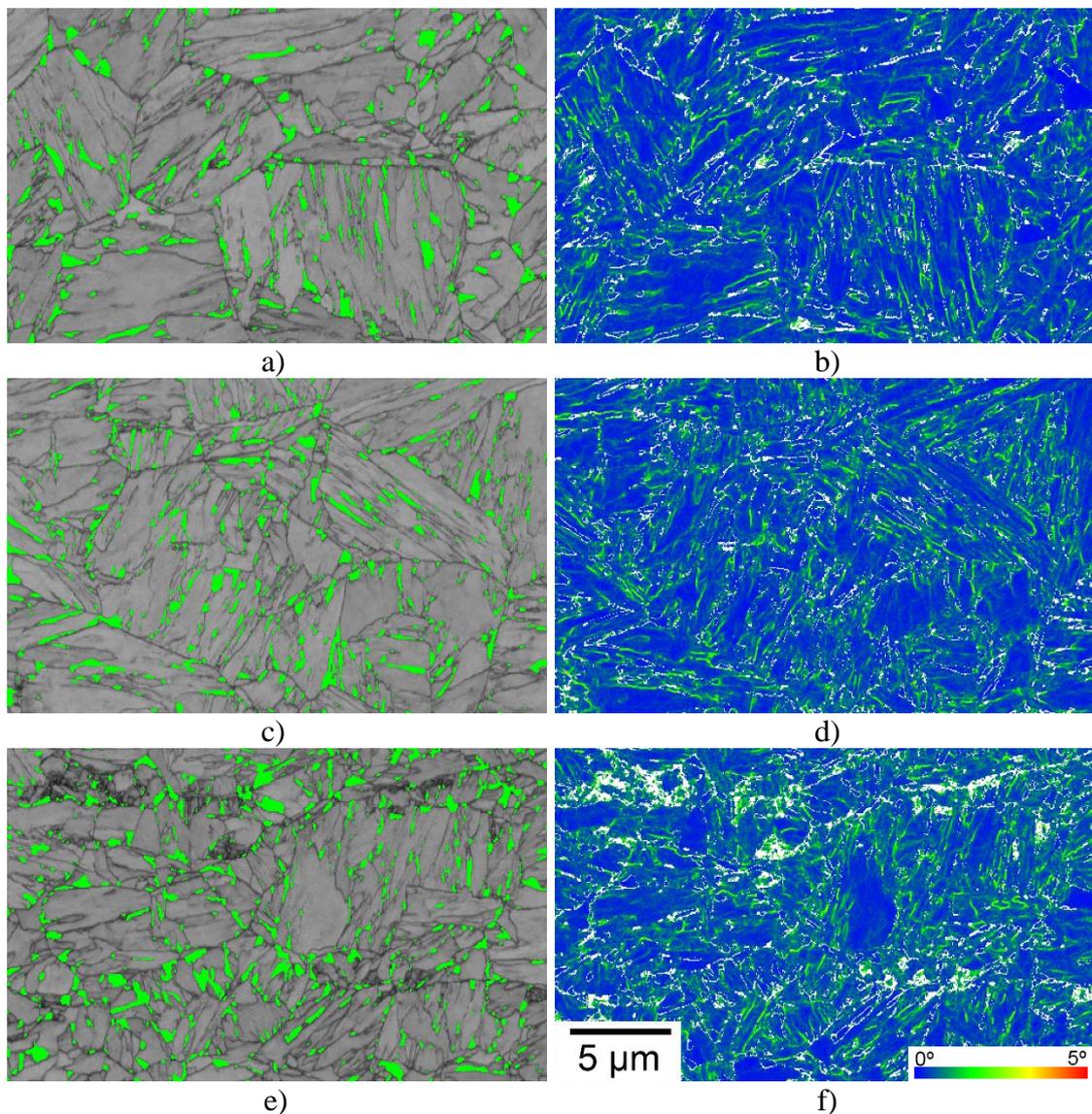

Fig. 7 Microstructure of the Q&P treated steels. a,c,e) Band contrast maps overlaid by phase map where retained austenite is in green; b,d,f) KAM maps; a,b) base steel; c,d) base+Cr steel; e,f) base+Cr+Nb steel. The scale bar 5 µm shown in (f) is valid for all maps.

The XRD measurements were additionally performed to measure the total volume fraction of retained austenite, and their outcomes are also provided in Table 6. It can be seen that the base+Cr steel shows a slightly higher fraction of retained austenite compared to the base steel, whereas the addition of Nb does have a noticeable effect, which is even more pronounced in the XRD results with respect to the ones derived from the EBSD analysis. Comparison of the EBSD and XRD data indicates that 2.9 – 3.3 % more retained austenite is present in the studied Q&P steels, which should be a film-type retained austenite, as it cannot be resolved in EBSD analysis due to its very small size (< 50 nm) [35].

Table 6 Data on microstructure of the Q&P steels before and after Nakajima testing. RA: retained austenite; FM: fresh martensite; PAGS: prior austenite grain size; $\varepsilon$: local true strain accumulated in the area of EBSD analysis.

| Steel | Initial (i.e. after Q&P treatment before Nakajima testing) | | | | | | After Nakajima testing | | | |
|---|---|---|---|---|---|---|---|---|---|---|
| | EBSD index rate [%] | RA by EBSD [%] | RA by XRD [%] | FM by EBSD [%] | PAGS [μm] | Martensite grain size [μm] | Sample type | EBSD index rate [%] | RA by EBSD [%] | $\varepsilon$ [%] |
| Base | 92.2 | 7.0 | 10.3 | 0.8 | 8.0 | 3.7 | A010 | 68.8 | 0.23 | 34.7 |
| | | | | | | | A050 | 83.5 | 0.63 | 28.1 |
| | | | | | | | A100 | 69.2 | 0.24 | 64.6 |
| Base+Cr | 92.7 | 8.0 | 10.9 | 1.1 | 8.3 | 3.6 | A010 | 73.4 | 0.22 | 27.4 |
| | | | | | | | A050 | 79.9 | 0.36 | 31.0 |
| | | | | | | | A100 | 67.8 | 0.21 | 65.8 |
| Base+Cr+Nb | 88.0 | 9.9 | 13.0 | 1.9 | 6.0 | 2.7 | A010 | 71.8 | 0.40 | 23.9 |
| | | | | | | | A050 | 81.7 | 0.54 | 18.0 |
| | | | | | | | A100 | 65.9 | 0.23 | 51.2 |

Analysis of prior austenite grain size via ARPGE software has revealed that alloying by Cr does not affect the average prior austenite grain size, which is 8.0 μm for the base steel and 8.3 μm for the base+Cr steel. However, additional microalloying by Nb noticeably reduces the prior austenite grain size to 6.1 μm, and it is also seen on the EBSD band contrast maps (Fig. 7a,c,e).

The latter effect can be related to suppression of austenite grain growth during austenitisation treatment due to pinning of grain boundaries by NbC nanoprecipitates. Indeed, TEM analysis of the base+Cr+Nb alloy revealed high amount of spherical NbC nanoprecipitates, homogeneously distributed in the material (Fig. 8). Two different types of NbC nanoparticles can be resolved on the TEM images: very fine having a size of 3 – 5 nm (marked by dashed arrows) and larger ones having a size of 10 – 15 nm (marked by solid arrows). Theoretically, a maximum 0.0032 wt.% of C out of 0.2 wt.% is consumed by the nominal Nb content for the formation of NbC (provided that all Nb atoms will form carbides). Therefore, the carbon loss due to NbC formation is negligible, and it does not negatively affect the volume fraction of retained austenite in the base+Cr+Nb grade, as seen from Table 6.

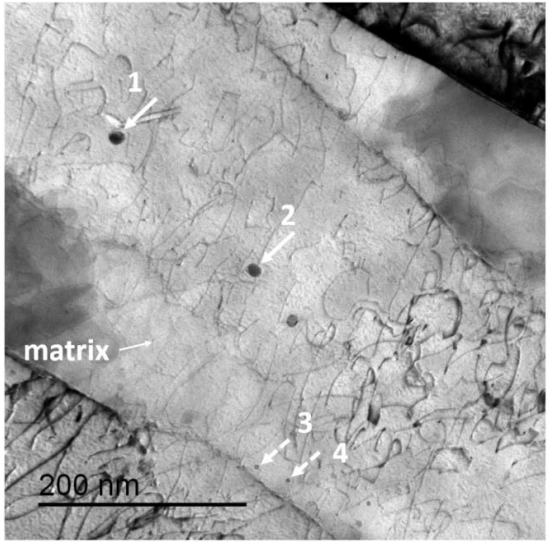

| Particle | Size [nm] | Nb | Cr | Si | Mn |
|---|---|---|---|---|---|
| 1 | 13.3 | 11.54 | 0.27 | 0.39 | 1.60 |
| 2 | 11.0 | 12.67 | 0.43 | 0.69 | 1.17 |
| 3 | 3.5 | 3.41 | 0.54 | 0.96 | 1.94 |
| 4 | 3.5 | 3.64 | 0.61 | 0.65 | 2.29 |
| matrix | - | 0 | 0.71 | 1.18 | 2.03 |

Fig. 8 TEM bright field image of the base+Cr+Nb steel after Q&P treatment. Very fine NbC nanoparticles having a size of 3 – 5 nm are marked by dashed arrows. Larger NbC nanoparticles having a size of 10 – 15 nm are marked by solid arrows. Table with data on chemical composition (in wt. %) of measured spots (marked on TEM image) is provided.

Two factors can be considered to enhance volume fraction of retained austenite in the base+Cr+Nb alloy. It is well known that Nb stabilizes retained austenite introducing a higher stacking fault energy [38]. However, EDX analysis of the matrix did not reveal any presence of Nb dissolved in the matrix (Fig. 8). Therefore, its contribution to the stabilization of retained austenite can be ruled out. The finer prior austenite grain size of the base+Cr+Nb alloy (Table 6) seems to be the main factor resulting in the enhanced volume fraction of retained austenite. Its effect on the microstructure evolution during Q&P process of carbon steels was thoroughly studied in [26]. Provided the sufficient volume fraction of primary martensite, and that the partitioning conditions ensure the partitioning of all carbon to the surrounding austenite, finer prior austenite grain sizes result in a faster and more efficient carbon partitioning process than coarser microstructures through the formation of smaller and more homogeneously distributed phases during the first quench.

*3.2.2 Effect of Nakajima testing on the microstructure of the Q&P steel grades*

Analysis of the microstructure of selected tested samples (geometries A010, A050 and A100 as shown in Fig. 4) was carried out on the RD-ND plane near the crack but out of the necking area. Theoretically, the geometry A010 presents the condition of uniaxial tensile deformation combined with plastic bending, the A050 sample is deformed in plane strain mode, whereas geometry A100 represents equi-biaxial stretching conditions. Fig. 9-11 illustrate the band contrast EBSD maps of the Q&P samples after Nakajima testing and relevant KAM maps. The results of quantitative analysis of the EBSD measurements are listed in Table 6. An information on the true local plastic strain $\varepsilon$ accumulated during Nakajima testing in the area analyzed by EBSD is also provided. The latter was estimated as $\varepsilon = \mathrm{Ln}\,(h_o/h_f)$, where $h_o$ is the initial thickness of the sheet and $h_f$ the final local thickness of the tested sheet in the area analyzed by EBSD. The following effects of Nakajima testing on the microstructure can be noted:

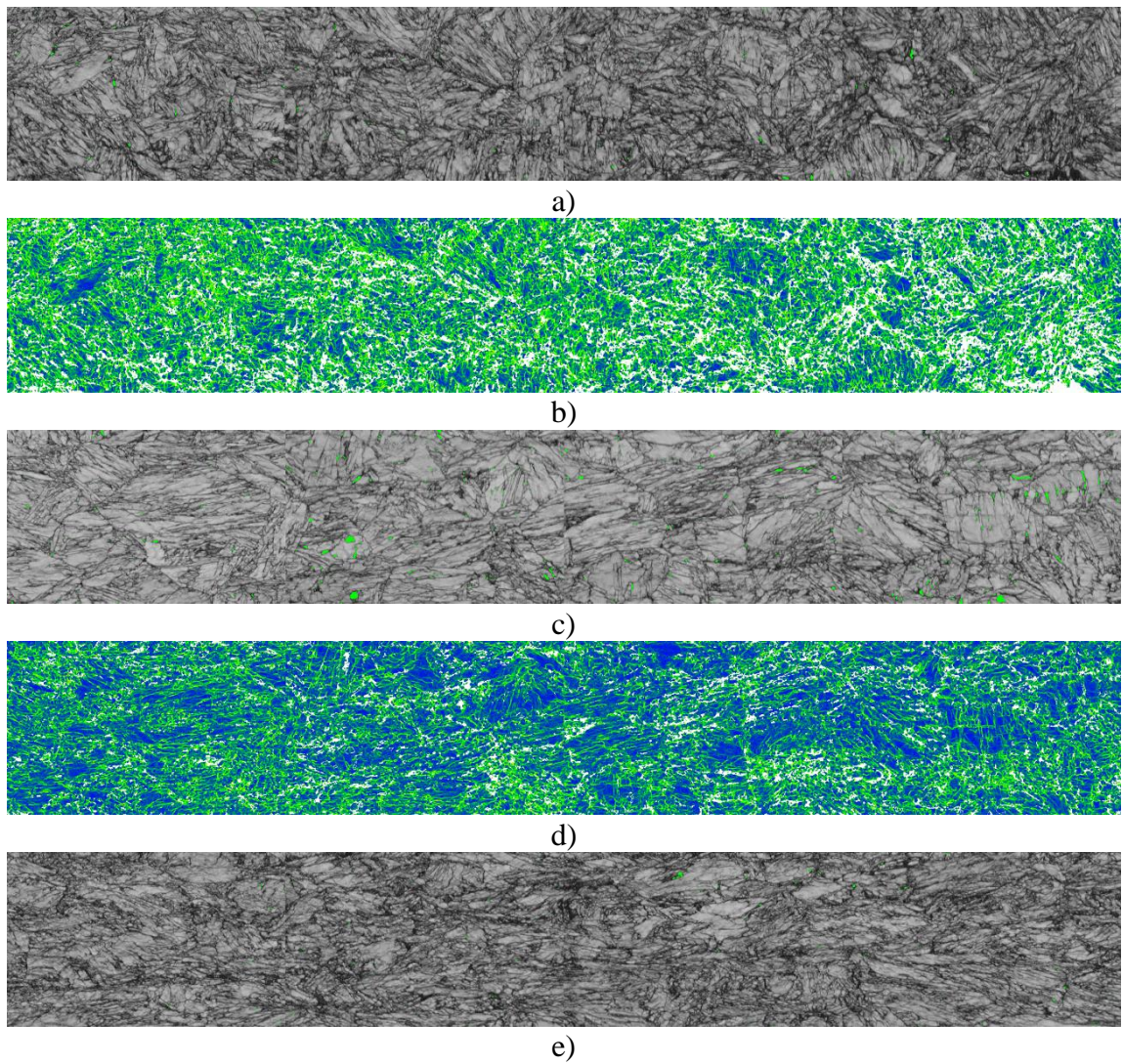

a)

b)

c)

d)

e)

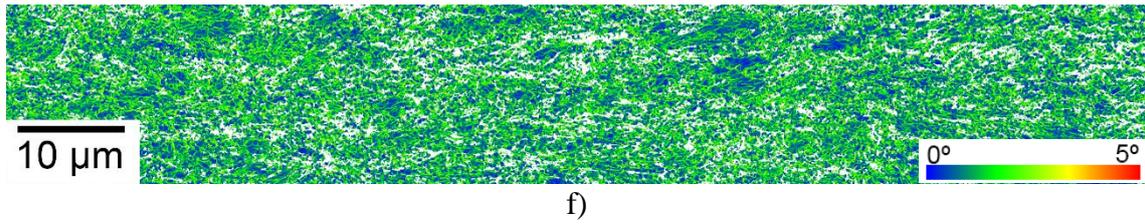
f)

Fig. 9 Band contrast maps with overlaid phase map for base steel samples after Nakajima testing: (a, b) A010 sample, (c, d) A050 sample, (e, f) A100 sample. The scale bar 10 μm shown in (f) is valid for all maps.

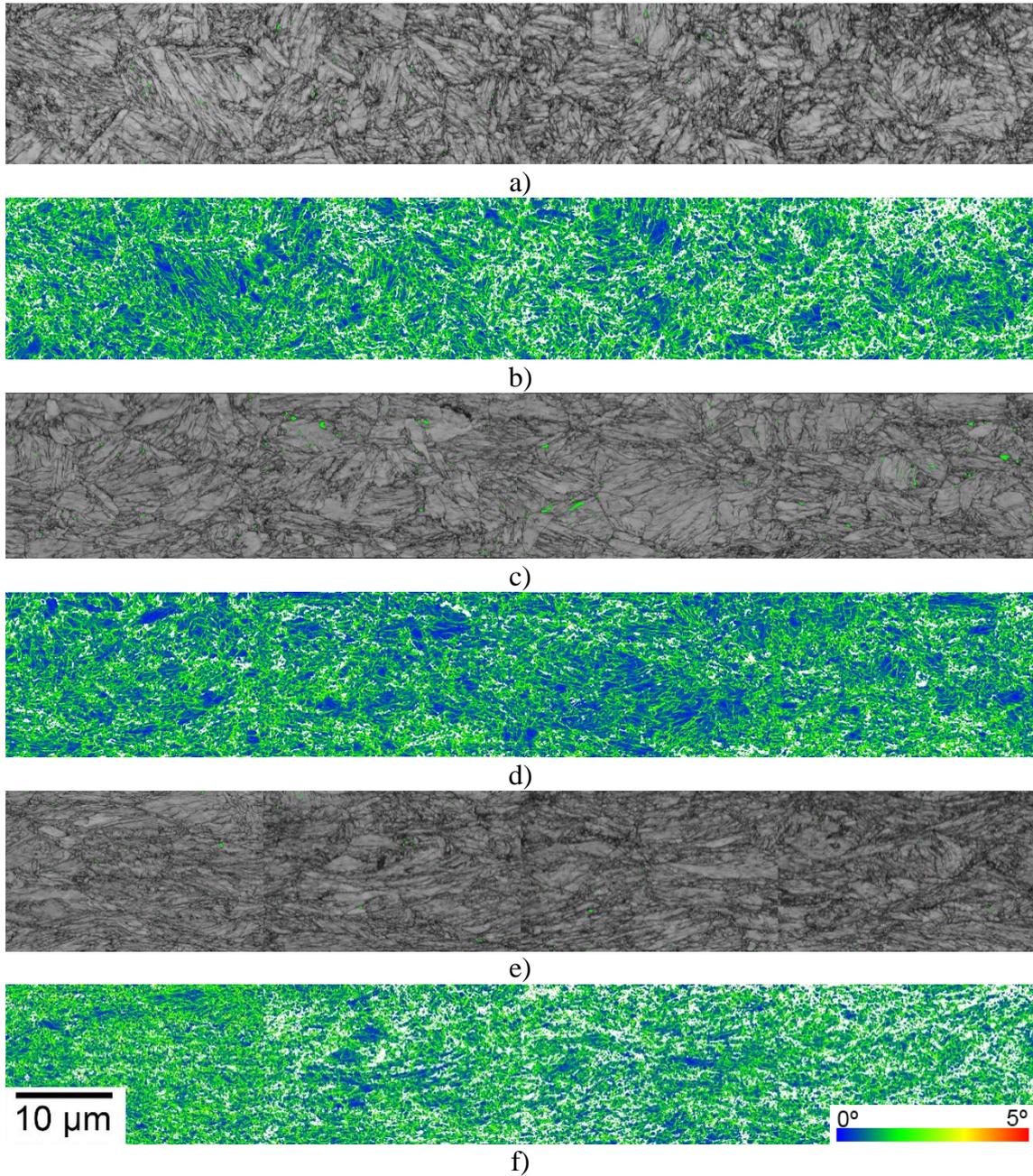

Fig. 10 Band contrast maps with overlaid phase map for base+Cr steel samples after Nakajima testing: (a, b) A010 sample, (c, d) A050 sample, (e, f) A100 sample. The scale bar 10 μm shown in (f) is valid for all maps.

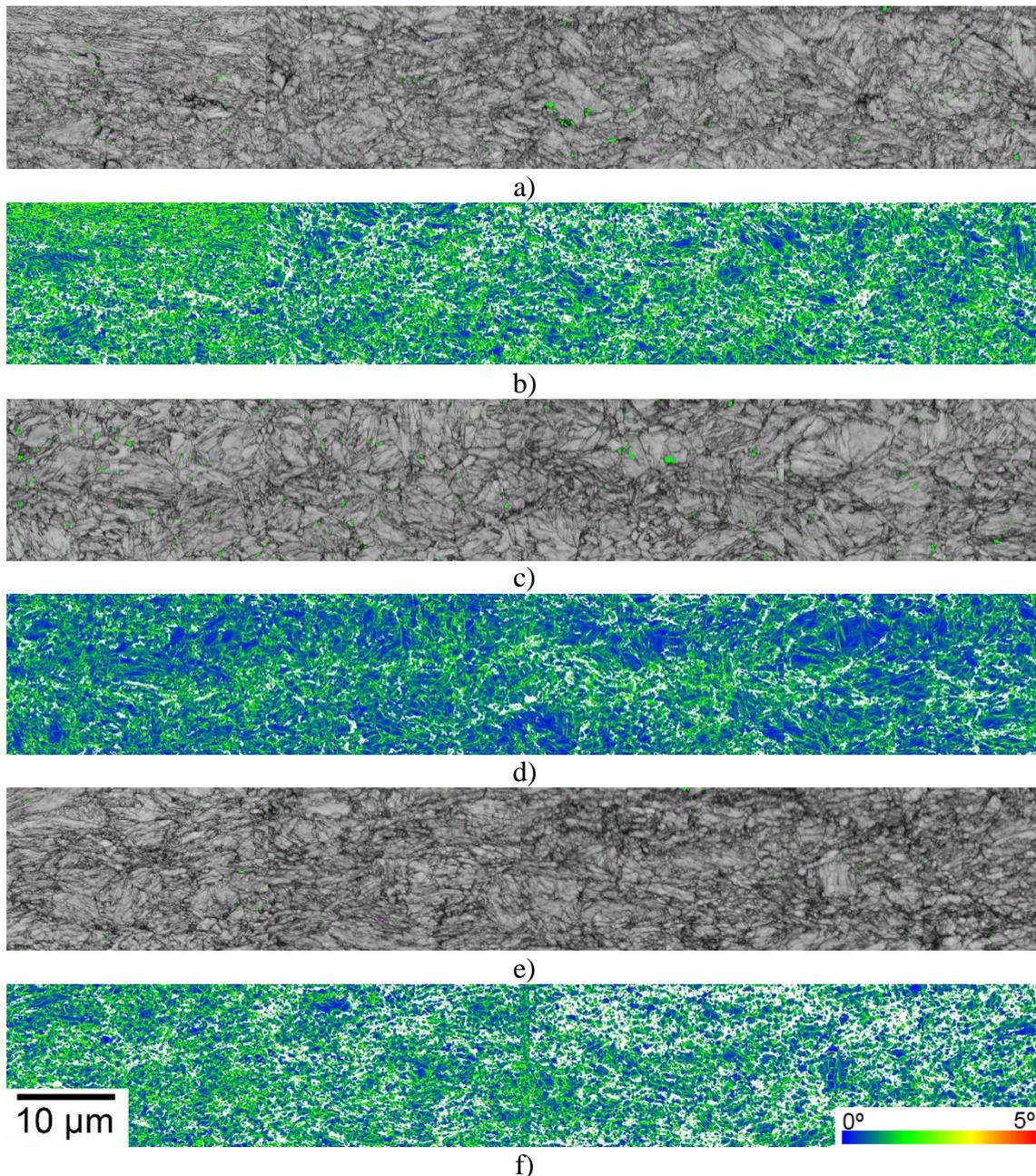

Fig. 11 Band contrast maps with overlaid phase map for base+Cr+Nb steel samples after Nakajima testing: (a, b) A010 sample, (c, d) A050 sample, (e, f) A100 sample. The scale bar 10 μm shown in (f) is valid for all maps.

1) The volume fraction of retained austenite dramatically decreases to values in the range of 0.21 – 0.63 % in all analyzed samples (Table 6). From band contrast EBSD maps (Fig. 9-11), it is clearly seen that mainly fine interlath austenite grains remain in the deformed microstructure. No clear correlation between sample geometry (i.e. amount of accumulated plastic strain in the area of EBSD analysis) and volume fraction of retained austenite is observed. For example, in the A010 sample of the base+Cr alloy (local plastic strain of 27.4 %) 0.22 % of retained austenite is present, whereas in the A100 sample (local plastic strain of 65.8 %) the volume fraction of retained austenite is 0.21 %.

2) Analysis of KAM maps shows that the local average misorientations have significantly increased after Nakajima testing. The histograms of misorientation distribution clearly demonstrate a shift towards higher values after Nakajima testing in all grades (Fig. 12a). It should also be noted that this shift is more pronounced in the samples that accumulated a higher amount of plastic strain, i.e. deformed in equi-biaxial stretching conditions (geometry A100) (Fig. 12b). Also the base A100 sample, which accumulated a higher amount of true plastic strain (64.6 %), shows a higher misorientation compared to the A100 base+Cr+Nb sample with a true plastic strain of 51.2 %. These higher misorientations are associated with the geometrically necessary dislocations (GNDs) [39]. Plastic deformation accumulated in the tempered martensitic matrix resulted in the formation of subgrain structure with the average size of ~0.5 µm, which is clearly seen on the KAM maps (Fig. 9-11).

3) In EBSD analysis, the Q&P treated base+Cr+Nb steel showed the lowest indexing rate of 88 %, whereas the Q&P treated base steel and the base+Cr steel showed higher and nearly similar values of 92.2 % and 92.7 %, respectively. The latter observation can be related to the fact that the base+Cr+Nb steel has the highest fraction of non-indexable fresh martensite (which appears as white pixels on KAM maps), as well as a finer microstructure (Table 6). The finer prior austenite grain size results in higher dislocation density, which should not significantly decrease after partitioning at 400 $^{o}$C for 50 s [40]. In the deformed samples, the indexing rate drops down to values in the range of 65.9 – 83.5 %, and a clear correlation between amount of introduced plastic strain and indexing rate is seen for each grade (Table 6). This observation is related to the formation of non-indexable stress/strain induced martensite during plastic deformation and the accumulation of a high amount of plastic strain in the interior of the tempered martensitic grains.

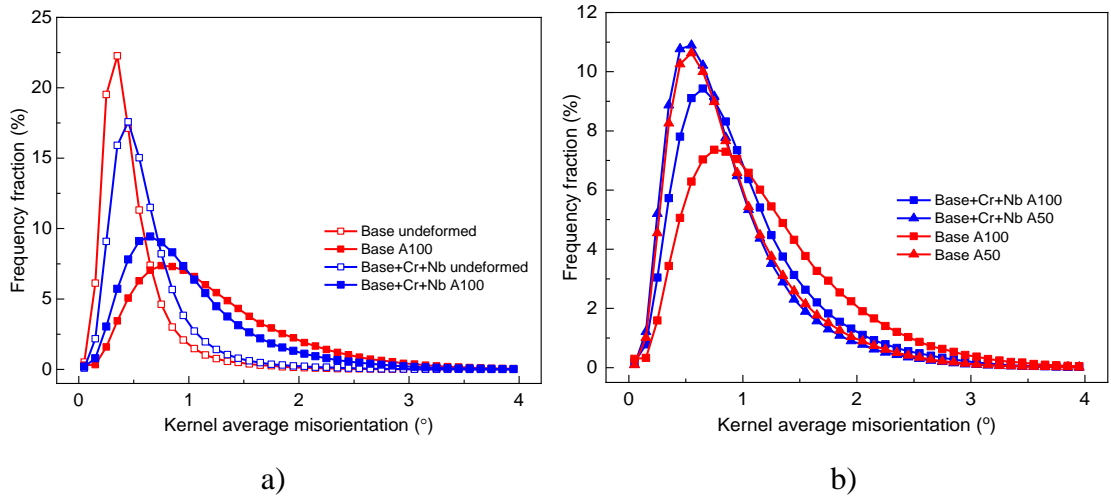

Fig. 12 a) The effect of Nakajima testing on the KAM distribution in selected samples (base and base+Cr+Nb); b) The effect of sample geometry (i.e. amount of accumulated plastic strain) on histograms of KAM distribution in selected samples (base and base+Cr+Nb).

## *3.3 Influence of microstructure on tensile behavior and formability of the Q&P processed steels*

The comparison of the experimental results clearly indicates that the ability of the studied Q&P grades to accumulate plastic deformation without cracking in Nakajima testing (Fig. 6, Table 5) is much higher compared to that in uniaxial tensile testing (Fig. 3, Table 6). This observation can be explained in terms of the presence of a strain gradient through the sheet thickness caused by the superposition of stretch-bending imposed by the punch and the die geometry inherent to the Nakajima test. This phenomenon, first addressed by Ghosh and Hecker [41], delays strain localization, allowing for higher levels of deformation before failure. This effect can be observed in conventional steels, but it is much more pronounced in AHSSs [42]. These materials exhibit a high hardening rate in the first stages of deformation (Fig. 3) compared to conventional steels, which would retard the onset of necking. A very important role of retained austenite should also be noted. As it is well known, austenite-martensite transformation during plastic deformation improves ductility/formability of AHSSs due to the TRIP effect [43,44]. One of the main factors determining the stability of retained austenite during plastic deformation is its crystallographic orientation with respect to the direction of the applied stress [43]. The dependence of the stability of austenite on the crystallographic orientation is defined by the theory of martensite crystallography. During uniaxial tensile deformation, the retained austenite grains with the high transformation potential transform into martensite, thus increasing the tensile ductility of the material, whereas the grains with low transformation potential remain in the microstructure [43]. In the case of multiaxial deformation (e.g. equi-biaxial) and presence of strain gradients through the sheet thickness, there are much less retained austenite grains with low transformation potential (i.e. which are stable against transformation). The vast majority of retained austenite grains transform into martensite thus increasing the formability of the AHSS, unless other factors (such as very small grain size or very high carbon content in the given grain) additionally stabilize an individual grain [43].

Another interesting observation is related to the controversial mechanical response of the base+Cr+Nb grade in the uniaxial tensile testing and Nakajima testing. In uniaxial tensile testing, the base+Cr+Nb grade shows the best strain hardening ability (Fig. 3), which also results in the highest average uniform elongation and elongation to failure values (Table 3). This observation could be related to the presence of NbC nanoprecipitates (Fig. 8) and the enhanced volume fraction of retained austenite compared to other studied grades (Table 6). NbC nanoprecipitates suppress grain growth resulting in finer PAGS (Table 6) which, in turn, positively affects ductility. Progressive transformation of retained austenite with plastic strain during tensile testing delays necking, that also leads to somewhat higher uniaxial tensile ductility of the material. However, in Nakajima testing, the base+Cr+Nb grade shows the lowest formability among all studied grades (Fig. 6, Table 6), though it is still much higher compared to the uniform elongation value demonstrated in uniaxial tensile test (Table 3). As it is mentioned above, the improvement in ductility in steels containing retained austenite is related not only to its volume fraction, but also to its mechanical stability. The more stable retained austenite in the steel progressively transforms during the whole straining process, resulting in higher work hardening rate which, in turn, delays the localization of plastic deformation. On the other hand, the less stable retained austenite grains transform at the earlier stage of plastic straining. Table 6 shows that the base+Cr+Nb A50 sample deformed to much lower amount of plastic strain (18.0 %) compared to the A100 sample (51.2 %) has nearly similar low volume fraction of retained austenite (0.23 – 0.54 %), which was 9.9 % before testing. Therefore, it can be hypothesized that the contribution of the phase transformation into plastic strain is negligible after reaching the true plastic strain of 18.0 % and other factors start playing an important role. A similar trend is also observed for the other two studied Q&P grades (Table 6). It should also be noted that a lower stability of retained austenite in the base+Cr+Nb steel compared to other grades was recently demonstrated in [28]. This was related to the lower retained austenite carbon content (1.08 wt.%) in the base+Cr+Nb steel than, for example, in the base grade (1.14 wt.%) and to its somewhat larger average grain size. Therefore, the retained austenite grains of the base+Cr+Nb grade during mutiaxial deformation may transform at the earlier stage of plastic deformation. The latter leads to the lower strain hardening ability of the material at the later stage of plastic deformation and earlier onset of necking and failure during Nakajima testing.

Apparently, while stress/strain induced martensite formation improves strain hardening during uniaxial straining it may deteriorate formability. What about other important factors to consider in the microstructural design of the Q&P steels to improve formability? The KAM maps presented in Fig. 9-11 and histograms of misorientation distribution (Fig. 12) clearly show that the ability of the tempered martensite matrix to accumulate plastic deformation is another important factor. It is determined by the chemical composition of the steel and the applied partitioning parameters (temperature and time). We have not studied the effect in the present work, as the same partitioning parameters were applied to all grades resulting in similar tempering conditions. However, an earlier study on the fracture behavior of Q&P steels has demonstrated that a better tempered martensitic matrix shows higher total crack growth resistance compared to harder counterparts [45].

**Conclusions**

The effects of alloying and microstructure on the formability of three advanced high strength steels (AHSSs) processed via quenching and partitioning (Q&P) was experimentally studied. It is shown that the uniaxial tensile ductility of the Q&P steels does not necessarily correlate with their formability. The Q&P steel with the highest volume fraction of retained austenite, i.e. the base+Cr+Nb steel, which showed the highest tensile ductility and strain hardening ability, demonstrated the lowest formability. The following explanations are proposed for the observed phenomenon.

- The low transformation stability of retained austenite in the material deformed under multiaxial loading with strain gradients through the sheet thickness leads to consumption of retained austenite at the earlier stages of plastic deformation, thus exhausting the material's strain hardening ability and resulting in an earlier necking and fracture of samples.

- The stress/strain induced martensite plays an important role in the formability of Q&P steels. The base+Cr+Nb Q&P steel with the highest volume fraction of retained austenite also contained slightly higher volume fraction of fresh martensite. Formation of microcracks at brittle fresh and stress/strain induced martensite at the earlier stages of plastic deformation may lead to their growth and failure of samples.
- Another important factor is the ability of the tempered martensitic matrix to accumulate plastic deformation during formability testing. It contributes to the strain hardening, thus delaying necking and fracture, as well as suppresses growth of microcracks formed in the material during formability testing.

**Acknowledgements**

The authors would like to acknowledge financial support by the Research Fund for Coal and Steel via OptiQPAP project (Grant Agreement 709755). Peikang Xia acknowledges gratefully the financial support from Chinese Scholarship Council (No. 201606890031).

**Data availability**

The raw/processed data required to reproduce these findings cannot be shared at this time due to technical or time limitations.